# The Impact of Cyber Security Threats on the 2020 US Elections

Nicholas Stedmon
Cyber Security and Human Factors
Bournemouth University
Poole, United Kingdom
i7826646@bournemouth.ac.uk

*Abstract*— This paper will investigate the literature surrounding cyber security threats in the 2020 US Elections. It begins with a brief overview of cyber security and the current state of cyber security regarding elections. In the main body of the paper, the focus will be on the literature review of three main areas: voter suppression, voter fraud, and disinformation, considering their impacts on the outcome of the election and on the voting public. Having evaluated sources on each this paper concludes by summarising the areas which have had the greatest impact on the 2020 US elections.

## I. Introduction

Cyber security has become a foremost concern in society; it protects all manner of data and systems, from personal emails to the controls of a nuclear power station. Naturally, it follows that the more we come to rely on technology, the more we rely on cyber security measures to protect us from external threats or internal vulnerabilities. Few events feel such tensions more keenly than an election, where tampering can subtly and significantly affect the outcome of the voting process, whether by altering the votes themselves, or discrediting candidates and voting measures [1].

The most prominent current example of cyber attacks threatening an election would be the United States of America, where the dangers of potential foreign interference in the process has been at the forefront of security concerns, particularly concerning Russia, though attribution is difficult to prove definitively [2] [3] [4]. In addition to the threat of experienced third-party attackers actively trying to cause harm, we must also consider the problems posed by the voting equipment. Some US voting machines were found to have substantial vulnerabilities built-in to them or not compensated for. Direct-recording electronic machines (DREs) used following the 2000 US elections were paperless; units installed to reduce issues with punch card ballots. As a result, they had no way of allowing a user to check their ballot before casting it, essentially leaving no paper trail (literally or figuratively) should there be an error or interference [3].

It is widely agreed that the use of physical ballots is the best way to prevent fallout from cyber attacks [1] and yet the Brennan Center for Justice projected that as many as 12% of the voting public would be using paperless voting machines [5]. The use of such outdated machines, coupled with the threat of external (or internal) attack, makes the question of cyber security an extremely pressing one.

Although the security of voting hardware is important, it's generally agreed that no ballots were altered in the 2016 attack [6], leading some to argue that the greater threat to the US elections is not so much the machines, but the infrastructure surrounding it [2]. In particular, the Cybersecurity and Infrastructure Security Agency (CISA) consider voter registration systems and pollbooks to be among the top elements to have the "greatest functional impact" [7] should they come under attack, a notion supported by evidence that data from voting registration systems was accessed in the 2016 election, possibly with the intent of using it for a future attack [2].

## II. Aims

In this paper I will review the existing research surrounding cyber security issues with US elections and consider the impact of such threats to the election process. Though cyber attacks can take a variety of forms, in the context of elections there are three key categories: voter fraud, voter suppression, and disinformation. Voter fraud is concerned with attacking or altering the outcome of an election; voter suppression involves discouraging individuals from casting their vote and/or disrupting the voting process; disinformation spreads false or misleading information about the voting process or candidates in order to colour voters' perceptions of the election.

## III. Literature Review

### A. Voter Suppression

In terms of voter suppression, two viable areas of attack are highlighted by the research: the voting infrastructure [1] and the machines themselves [8]. Part of the danger comes from election officials suffering an "information gap" [1, p. 61] in their understanding of online threats, meaning that the individuals writing security policy sometimes do not know how best to counter the problem. Hoke [1] suggests this deficiency in comprehension, coupled with a lack of appropriate safeguarding, is the shatter point for election cyber security. Another factor which compounds this problem is the diversity of the electoral ecosystem in the United States, with elections being managed "at the state and local levels, with a limited federal role" [9, p. 8].

In 2016 a White House Directive "put civilian agencies – not the Department of Defense – at the forefront of managing a cyber attack" [4, p. 32]. Pope [4] contends this directive, while acknowledging the "hybrid nature" [4, p. 32] of cyber attacks, is insufficient with addressing the issue and only led to a greater lack of clarity on the procedure of dealing with an attack. This climate of incomprehension has led to the formation of volunteer groups such as Election Cyber Surge (ECS) who work to uncover weaknesses in cyber security in the name of ensuring a safe and secure voting process [10]. CISA also offers its services to evaluate election cyber security, though officials often find it difficult to make time for such a review during election season [9]. Though concerned largely with supply chain risk management,

Hodgson et al [9] do investigate the wider implications of a supply chain attack and note that this lack of a cohesive infrastructure means that protecting against cyber threats is much more difficult, since there is little consistency between different regions. This indicates that cyber security is an issue which needs to be addressed at the top, in order to prevent vulnerabilities trickling down throughout the whole election process. Pope [4] takes this idea further and suggests that there should be a greater focus on interagency collaboration, given that cyber criminals are often "transcending borders" [4, p. 27].

Outdated equipment is another vulnerability, one inherent in elections due to their infrequent occurrence, as voting hardware, software, and databases face periods of disuse and are therefore updated and maintained less often than other electronic systems [8]. Even if an attacker is not able to gain access to a machine to alter votes, the outdated systems could be more vulnerable to attacks that disrupt its functionality.

Furthermore, the "three largest vendors of U.S. election equipment" [9, p. 1] all use parts which have been sourced from Russia and China rather than the United States, and utilise many of the same components, meaning that a single attack could cause exponential problems throughout the supply chain.

Replacing the outdated machines with newer ones has the benefit of reducing known vulnerabilities but there is unintentional negative consequence in replacing DRE machines with optical scan machines – voting machines which accept ballots inserted into them and then scan the ballot to record the vote. Although optical scan machines allow for ballot auditing, reducing the risk of a cyber attack going unnoticed [1], units were susceptible to clogging if a ballot was dampened with hand sanitiser [11]. Due to the ongoing coronavirus pandemic, most polling stations were equipped with hand sanitisers to prevent the spread of the virus, which meant that voters whose hands were still wet with sanitising gel could inadvertently clog the machine or spoil their ballot. The problem had been noted previously, as a case in Bristol, VA in 2009 resulted in a machine being inoperable once it had received a damp ballot [12] but not compensated for, creating a situation where the machines which were intended to decrease the likelihood of incorrect votes, may have resulted in some individuals' votes not being counted at all.

A public service announcement from the FBI and CISA discount claims election systems could be disrupted or changed in order to invalidate votes, stating that even if attackers held voter information, it would not prevent citizens from voting or be able to alter the vote [13]. They did concede, however, that such attacks could delay the voting process by "render(ing) these systems temporarily inaccessible" [14]. Any delay in accessing the voting machines could result in some voters being unable to submit their ballot before the polling stations close – if there are not enough alternative machines available - or equally could delay election officials from counting the votes contained within the units. While the machines represent a physical vulnerability in the election system, it is one that would be improved by tighter policy controls and more frequent testing of the existing units.

*B. Voter Fraud*

Voter fraud is a potential threat which is compounded by the advent of electronic voting machines [1] to the point where paper ballots remain the "gold standard" [5, p. 18] when it comes to preventing vote tampering. However, despite the significant threat posed by electronic voter fraud, there has been little research into this subject when compared with suppression and disinformation – a number of papers dealing with voter fraud do not even include the words 'cyber' or 'online' in relation to voter fraud [15], [6] [16]. Kiyohara [17] writes briefly on the subject of electronic fraud, though only as part of a wider study on online voter registration (OVR). She notes that Republicans are more likely than Democrats to express concern over the potential of fraud in OVR practices [17], a tendency other sources also reflect [16] [18].

Much of the literature surrounding voter fraud focuses exclusively on traditional means. It is possible that this is due to the "dearth of evidence" [6, p. 123] of significant levels of fraud, but given the previously established pressure points in the US' election infrastructure, the potential for voter fraud cannot be entirely disregarded. In addition, even if there is no proof of fraud, such allegations can affect public opinion [15] which leads into the topic of disinformation.

*C. Voter Disinformation*

A somewhat nebulous, but perhaps widest reaching, impact of cyber attacks is how they colour public perception. The spread of disinformation is something of a grey area in terms of cyber security, as there may not be any 'attack' taking place in the literal sense, but the misleading or false information must nonetheless be removed. Doing so will often require "interagency collaboration" [8, p. 6] and usually relies on the companies controlling the media platforms to step-in or tighten their terms of use. Many of the issues which exacerbate or mitigate voter suppression are also applicable to disinformation campaigns. The diversity of the electorate across the US means that any disinformation campaign will prey upon the most susceptible demographics to increase division [19]. Another similarity with voter suppression is in terms of response; to prevent widespread confusion, a strong and swift government response is needed to quickly dispel false information [19]. Pope [4] notes that any response to disinformation must be a delicate balancing act between "counter(ing) the threat…while avoiding any influence on the outcome of the elections" [4, pp. 25-26], hence why the response to allegations of Russian interference in 2016 was somewhat muted. Hansen and Lim [19] argue that governments' cyber security focus is driven by military concerns and thus less interested devoting resources to the social ramifications. Election officials' limited knowledge on the subject [1] seems to give tacit support to this theory.

A Gallup poll from 2019, similarly indicated general distrust in the election process, with 59% of those polled did not believe the results of the election would be honest, polling lower than much of Europe and substantially lower than the Nordic countries polled [20].

Whether real or imagined, and regardless of the source, the climate of fear of cyber attacks has shaken many Americans' confidence in the voting process, as well as the results [3]. For some, even producing evidence that the chance of interference of fraud is low is not sufficient to prove the security of the election and "the meme of voter fraud" [21, p. 760] is still perpetuated. In a Marist Poll [18], 47% of Republicans believed that voter fraud was the greatest risk to a "safe and accurate" [18, p. 11] election. Just over half of polled voters believed that President Trump was actually encouraging election interference, and in terms of the sample break down, very few groups strongly believed he was making the election safer. It is prudent at this juncture to note that the poll question only offered three response options to this topic, with the third being "unsure" [18, p. 10]. Therefore, the rather binary results may belie a more nuanced range of opinions; for example, it does not account for someone believing Trump has had no effect, positive or negative, on the process, or that he has made the election less safe, but not necessarily encouraged interference.

On the flip side of this coin, the same poll indicated that Democrats believed voter suppression would be the greatest risk (34%), while Independent voters believed misleading information was (39%), the latter risk being the highest overall at 35% of all polled, with voter fraud second at 24%, and voter suppression third at 16% [18].

As the events of November 2020 unfolded, the true impact of disinformation became patently clear: no matter the proof or veracity, if members of the public do not believe the voting process to be reliable, "the actual accuracy (of the)… electoral results mean little" [5, p. 25].

## IV. CONCLUSION

Despite significant concern over the possibility of a breach in cyber security, the US 2020 election was ultimately unaffected by direct cyber attacks, in terms the election's outcome or ballot tampering. The cyber security measures in place were sufficient to block or deter attackers from affecting the results or the voting process. Nonetheless, the volume of disinformation online and the climate of unease posed by cyber attacks has had a considerable effect. While the literature may have differing views on the most pressing issues, it is clear that the consensus is that the United States remains underprepared for, or at least underappreciative of, the dangers of cyber threats. The combination of ill-informed officials, widespread disinformation, and poorly maintained or outdated equipment has left the US election process vulnerable and without a more centralised and rigorous approach to cyber security, the United States remains an inviting target.

## REFERENCES

[1]  C. Hoke, "Internet voting: structural governance principles for election cyber security in democratic nations," in *GTIP '10: Proceedings of the 2010 workshop on governance of technology, information and policies*, New York, 2010.

[2]  S. Malempati, "The elections clause obligates congress to enact a federal plan to secure U.S. elections against foreign cyberattacks," *Emory Law Journal,* vol. 70, no. Forthcoming, 2020.

[3]  N. Inkster, "Information warfare and the US presidential election," *Survival,* vol. 58, no. 5, pp. 23-32, 2016.

[4]  A. Pope, "Cyber-securing our elections," *Journal of Cyber Policy,* vol. 3, no. 1, pp. 24-38, 2018.

[5]  K. Breedon and A. Bryant, "Conflicts of interest and election cybersecurity: how bipartisan congressional oversight can inform the public, address election system vulnerabilities, and increase voter confidence in election integrity," *Wayne Law Review,* vol. 66, no. 1, pp. 13-62, 2020.

[6]  D. Cottrell, M. C. Herron and S. J. Westwood, "An explorartion of Donald Trump's allegations of massive voter fraud in the 2016 general election," *Electoral Studies,* vol. 15, no. Feb 2018, pp. 123-142, 2018.

[7]  Cybersecurity and Infrastructure Security Agency, "Election infrastructure cyber risk assessment," 28 07 2020. [Online]. Available: https://www.cisa.gov/sites/default/files/publications/cisa-election-infrastructure-cyber-risk-assessment_508.pdf. [Accessed 16 11 2020].

[8]  S. Van der Staak and P. Wolf, "Cybersecurity in elections," International Institute for Democracy and Electoral Assistance, Stockholm, 2019.

[9]  Q. Hodgson, M. Brauner and E. Chan, "Securing US elections against cyber threats: considerations for supply chain risk management," RAND Corporation, 2020.

[10] Harris Public Policy, "Election cyber surge," University of Chicago, 2020. [Online]. Available: https://harris.uchicago.edu/research-impact/centers-institutes/cyber-policy-initiative/election-cyber-surge. [Accessed 25 11 2020].

[11] Centers for Disease Control and Prevention, "Polling locations and voters," 2020. [Online]. Available: https://www.cdc.gov/coronavirus/2019-ncov/community/election-polling-locations.html. [Accessed 03 11 2020].

[12] P. Limburg, "Optical scan/hand sanitizer warning!," ca. 2009. [Online]. Available: https://www.eac.gov/sites/default/files/eac_assets/1/28/Optical%20Scan.pdf. [Accessed 03 11 2020].

[13] Federal Bureau of Investigation, "Safeguarding your vote: A joint message on election security," 06 10 2020. [Online]. Available: https://www.fbi.gov/video-repository/interagency-election-security-psa-100520.mp4/view. [Accessed 02 11 2020].

[14] Internet Crime Complaint Center IC3, "Cyber threats to voting processes could slow but not prevent voting," 24 09 2020. [Online]. Available: https://www.ic3.gov/Media/Y2020/PSA200924. [Accessed 29 10 2020].

[15] J. Levitt, "The truth about voter fraud," 2007.

[16] M. Levy, "Winning cures everything? Beliefs about voter fraud, voter confidence, and the 2016 election," *Electoral Studies,* in press.

[17] S. Kiyohara, "Adoption of online voter registration systems as the new trend of US voter registration reform," *The Japanese Journal of American Studies,* no. 31, pp. 31-51, 2019.

[18] Marist College Institute for Public Opinion, "1/21 Election security," 21 01 2020. [Online]. Available: http://maristpoll.marist.edu/wp-content/uploads/2020/01/NPR_PBS-NewsHour_Marist-Poll_USA-NOS-and-Tables_Election-Security_2001140949.pdf#page=3. [Accessed 02 11 2020].

[19] I. Hansen and D. Lim, "Doxing democracy: influencing elections via cyber voter interference," *Contemporary Politics,* vol. 25, no. 2, pp. 150-171, 2019.

[20] R. J. Reinhart, "Faith in elections in relatively short supply in U.S.," Gallup, 13 02 2020. [Online]. Available: https://news.gallup.com/poll/285608/faith-elections-relatively-short-supply.aspx. [Accessed 29 10 2020].

[21] A. R. Ellis, "Voter fraud as an epistemic crisis for the right to vote," *Mercer Law Review,* vol. 71, no. 3, pp. 757-778, 2020.